\documentclass[11pt]{article}
\usepackage{amssymb,amsmath, amsthm}
\usepackage{graphicx,picinpar,epsf}
\usepackage{float}
\usepackage[margin=2.5cm]{geometry}

\usepackage{mathrsfs}
\usepackage{color}
\usepackage{soul}
\usepackage{algpseudocode}
\usepackage{algorithm}
\usepackage{amsmath}
\usepackage{bbm}
\usepackage{url}
\usepackage{subfigure,adjustbox}
\begin{document}
\title{Complexity of Possibly-gapped Histogram and \\
Analysis of Histogram (ANOHT)}
\author{\normalsize Hsieh Fushing\footnote{Correspondence: Hsieh Fushing, University of California at Davis, CA
95616. E-mail: fhsieh@ucdavis.edu} \hspace{1mm} and Tania Roy \\
Department of Statistics, University of California, Davis}
\date{\today}

\maketitle

\section*{Abstract}
Through real examples, we demonstrate that gaps and distributional patterns embedded within real-valued measurements are inseparable biological and mechanistic information contents of the system. To discover such patterns, we develop data-driven algorithms to construct possibly-gapped histograms. Then, based on these computed patterns, we propose a geometry-based Analysis of Histogram (ANOHT) to provide multiscale comparisons among multiple treatments. Our developments begin by showing that constructing a possibly-gapped histogram is a complex problem of statistical mechanics. Since its ensemble of candidate histograms is described via a two-layer Ising model in order to accommodate all local and scale-sensitive boundary parameters. This histogram construction is also a distinctive problem of Information Theory. Since, from the perspective of data compression via Uniformity, the decoding error criterion is shown as nearly independent of sample size. So if a Hamiltonian is defined as the sum of total coding lengths of boundaries and total decoding errors within bins, then the direct search for the macroscopic state or optimal solution is computationally impossible. We resolve this computing issue by applying the Hierarchical Clustering algorithm to narrow the search within a small set of nearly optimal solutions. An algorithm is then developed to select the best one. Further, based on a computed possibly-gapped histogram, the first phase of ANOHT is developed for simultaneous comparison of multiple treatments locally and globally, while the second phase of ANOHT is to develop a tree-geometry and to check the authenticity of branch-induced treatment-grouping by applying classic empirical process theory. The well-known Iris data is used to illustrate our technical developments, and a large baseball pitching data set and a heavily right-censored divorce data are analyzed to showcase the existential gaps and utilities of ANOHT.

\noindent
{\bf Keywords:} Data Mechanics; Hierarchical Clustering Algorithm; Macrostate; Statistical Mechanics; Unsupervised Machine Learning.
%\newpage
\section{Introduction}
Without spatial and temporal coordinates, a sample of one-dimensional (1D) real-valued measurements is generally taken as one basic simple data type and receives very limited research attention.  Its simplicity is seemingly implied by the illusion that its information contents are evident and transparent, and all its characteristic patterns should have immediately popped up right in front of our eyes. As a matter of fact, its pictorial representation, usually called an empirical distribution, indeed embraces hidden and implicit patterns waiting to be extracted.

There are many possible patterns that can be exhibited through the piecewise step-function structure of an empirical distribution. Among all possible patterns, two of them take the most basic forms: one is ``linear segment'' and the other is ``gap''.  A linear segment indicates a potential Uniform distribution being embedded within the empirical distribution, and a gap strictly indicates an interval zone, in which definitely allows no observations. As for the rest of the potential patterns, they can be very well approximated by properly combining these two basic patterns. Therefore, an empirical distribution ideally can be well approximated by various serial compositions of basic patterns. Each composition is a possibly-gapped piecewise linear approximation, which is correspondingly equivalent to a possibly-gapped histogram.

From such an approximation perspective, the larger number of patterns involving in such a series of basic patterns means a higher cost in terms of data compression \cite{Cover}.  Nevertheless, for the sake of true and intrinsic data patterns, it seems very natural to think that the hypothetical population distribution underlying the observed empirical distribution should embrace discontinuity.

But this is not the case in the statistical literature. Even though a histogram is discrete in nature, most existing versions of its constructions were developed through approximating a density function by imposing continuity and smoothness assumptions \cite{hall}, \cite{Rissanen}, \cite{Yu1992}. In sharp contrast, in computer science literature, a histogram is always discrete because it is primarily used for visualizing data or queries from a database \cite{Kontkanen}. So it does not involve with approximations via continuous components of Uniform or other smooth kernel distributions.

The computing costs for a histogram via aforementioned Statistics and Computer science viewpoints are not severely high. But the computing load for a possibly-gapped histogram can be theoretically very heavy because a constructive approach ideally needs to allow an unknown number of bins with heterogeneous bin-width and to accommodate an unknown number of gaps of different sizes. Further, the computing cost will grow with the sample sizes because of the multi-scale nature of these two basic patterns. That is why the construction of a possibly-gapped histogram  is computationally complex.

In this paper, we resolve this computational issue algorithmically by treating it as if it is derived from a physical system. From this perspective, a possibly-gapped histogram should embrace deterministic structures, which are all boundaries of the bins, and the stochastic structures that are Uniform within each bin. That is, the deterministic and stochastic structures together constitutes the system information contents of a one-dimensional data set.

Such physical information contents render another interpretation from the perspective of algorithmic complexity \cite{Li}. A relatively simple way of seeing this complexity is the fact that the candidate ensemble is constructed via a Two-layer Ising model \cite{Ising}. This ensemble grows exponentially in size with respect to the number of data points, say $n$. Based on this ensemble description, we clearly see that all boundary parameters are characteristically local because of multiple relevant scales also depending on $n$ in a heterogeneous fashion across all potential bins.

Further, it is interesting to note that the decoding error under Uniformity is nearly independent of sample size within each bin. This fact becomes an effective criterion for confirming Uniformity, on one hand. On the other hand, any further division of a Uniform distribution would give rise to several Uniform distributions with lower total decoding errors. Therefore we need to balance between the total decoding errors and coding lengths of boundaries of the bins (with respect to a converting index). So a Hamiltonian is defined.

Therefore constructing a possibly-gapped histogram is indeed a complex physical problem of statistical mechanics aiming for extracting the lowest Hamiltonian macroscopic state (or macro-state). It is also a problem of information theory aiming for the best balance between the costs. But it is clearly not a problem of statistics because of its multi-scale nature. Thus we need a brand-new computing protocol to seek for the optimal solution and the macro-state.

It is surprising that this seemingly complex macro-state can be approximated by a relatively simple computational algorithm. By applying the popular hierarchical clustering (HC) algorithm with complete or other modules, excluding the single-linkage one, on a one-dimensional dataset, the resultant hierarchical tree indeed gives rise to just a few feasible candidates. We develop an algorithm to select one from this small set of candidates. Nearly optimal solutions can be derived. Also, the criterion of decoding error of Uniformity turns out to be a practical way of checking whether a space between two consecutive bins is indeed a gap.

The merits of a possibly-gapped histogram are intriguingly profound and far-reaching. We demonstrate its potential merits through our developments of new data analysis paradigm, called Analysis of Histogram (ANOHT). The first phase of ANOHT is designed to address the issues: where and how multiple treatments or distributions are locally and globally different? The second phase of ANOHT is designed to answer issues: which treatments are closer to which, but far away from others? And why? Here ANOHT is simple and equipped with excellent visualization capability. ANOHT also gives the biological and mechanistic meanings to identify gaps found within a histogram.

Our technical developments are illustrated throughout by employing the well-known Iris data. In the Result sections, ANOHT is first applied onto a large pitching data set of three Major League Baseball (MLB) pitchers, and then onto a heavily censored divorce data. The implications of our developments from building a possibly-gapped histogram to ANOHT are discussed in the Conclusion section.

\section{Methods}
In this section, our technical developments are divided into four subsections: 1) building the ensemble of candidate histograms; 2) deriving the decoding errors; 3) constructing the algorithm for possibly-gapped histogram as the first phase of ANOHT; 4) developing the tree geometry as the second phase of ANOHT. The Iris data is illustrated throughout our developments.

\subsection{Candidate ensemble.}
\begin{figure}
\begin{center}

\includegraphics[height=2in, width=3.2in]{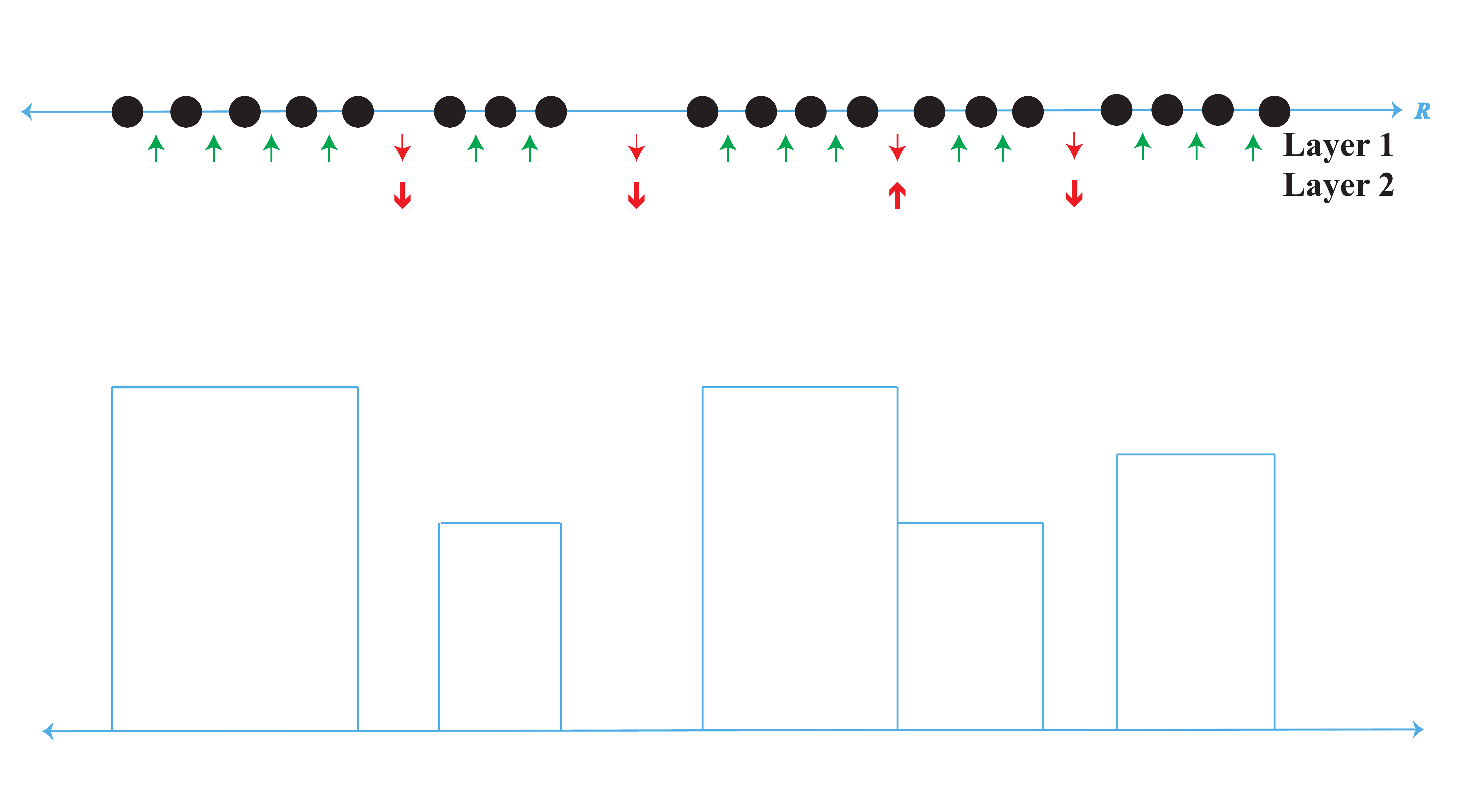}

\end{center}
\caption{Schematic illustration of two-layer Ising model.  The black dots on the blue line represent observations ordered and placed on the Real line.
The two layers of spins are given below in the next two rows, where up and down arrows describes``+'' and ``-'' spins respectively.
A green arrow means that the two consecutive pairs of observations will be in the same bin and a red arrow means they will be in different bins.
Hence, in the second layer, the down red arrow indicates two separate bins with an existential gap, and an  upward red arrow indicates two consecutive bins, but with no gap. Next, the histograms are constructed accordingly.}
\label{fig:Ising}
\end{figure}

Let $\{x^o_i\}^n_{i=1}$ be the observed data points, and their ranked values as $\{x^o_{(i)}\}^n_{i=1}$ in increasing order. The construction of a possibly-gapped histogram can be precisely stated in the following Two-layer one-dimensional Ising models depicted in Fig. \ref{fig:Ising}.

 \begin{description}
\item[1]Layer-1: a ``spin'' is placed on each spacing between a pair of consecutive observed values $(x^o_{(i-1)}, x^o_{(i)})$. First an Up-spin for sharing the same Uniform-part, and then a Down-spin for indicating that $(x^o_{(i-1)}$ , $x^o_{(i)})$ belongs to two distinct bins with or without a gap;
\item[2]Layer-2: a ``hidden-spin'' is placed between two consecutive Uniform-parts, on the first layer: 1) ``+''-spin for having a gap, and 2) ``-''-spin for without a gap.
\end{description}

\begin{figure*}
\includegraphics[height=6in, width=7in]{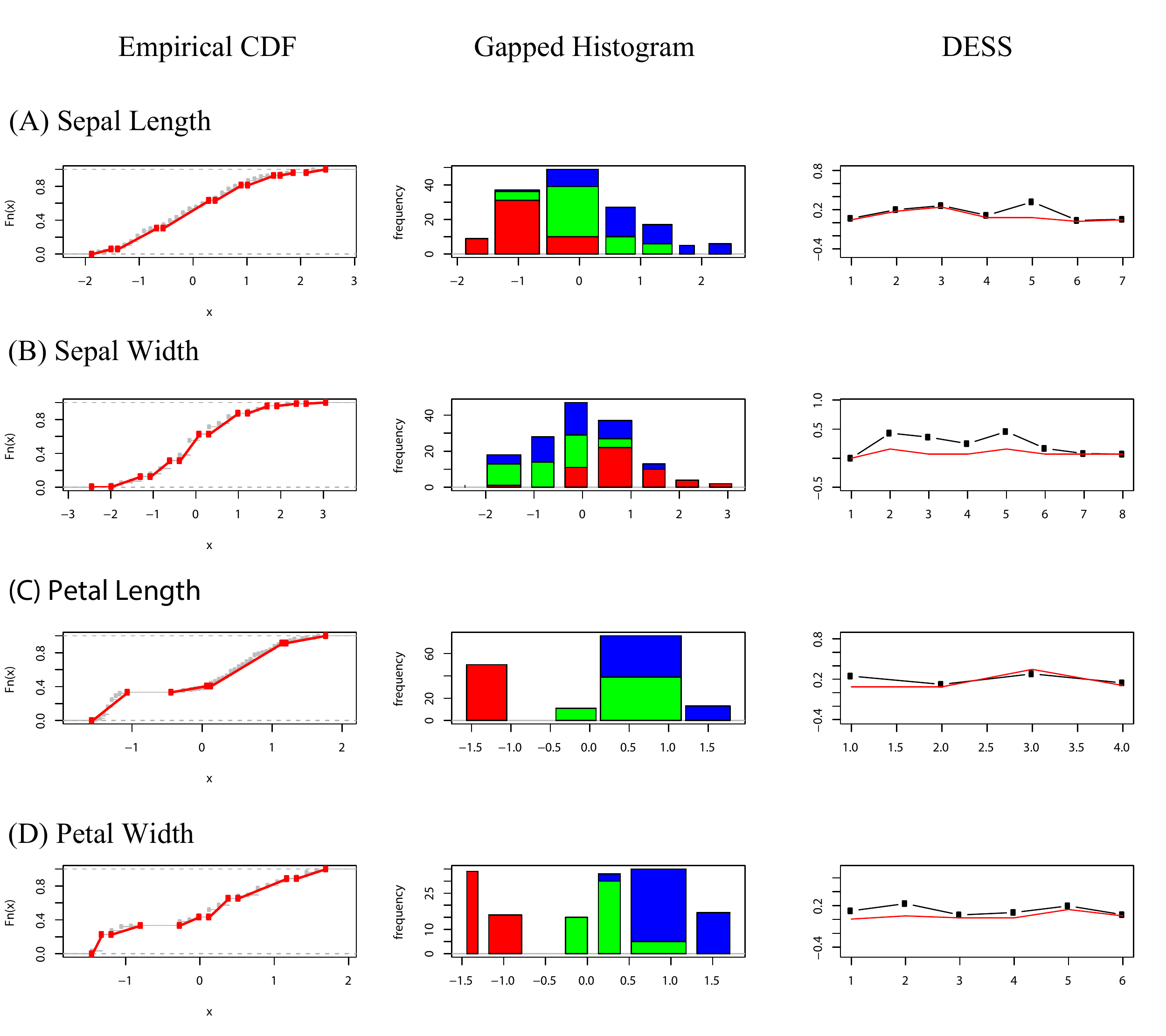}
\caption{The gapped histograms of the four columns of Iris dataset (namely, petal  length, petal width, sepal length and sepal width respectively) are displayed in four rows above. The proportion of  three species Setosa, Virginica, and Versicolor in each histogram is respectively represented by red, green and blue. The left column of panels shows the corresponding empirical cumulative distribution function, separated by red lines indicating each different bin from the histogram. The middle column of panels show the four possibly-gapped histograms via the hierarchical clustering algorithm with ``complete" module and index $L_0= 0.1 \times$(tree height). The right column of panels describes the DESS for each bin of the histograms, along with the threshold $(b_j-a_j)^2/3$ plotted in red color.}
\label{fig:hist_iris}
\end{figure*}

In total there are $n-1$ Up or Down-spins in the Layer-1. Then the number of spins in the Layer-2 is exactly the number Down-spins, say $k$, in the Layer-1.   Since $C^{n-1}_{k}$ is the number of distinct sets of $k$ nodes chosen from the pool of $n-1$ nodes, this class of ``Two-layer one dimensional Ising model'' has its cardinality growing exponentially with the sample size as:
\[
3^{n-1}=\sum^{n-1}_{k=0} C^{n-1}_{k} 2^k.
\]
This exponential growth rate reflects the essence of the fact that all boundaries are local once we revoke continuity and smoothness assumptions. That is, the complexity of computing within this ensemble comes from dealing with multi-scales of parameters, which is one significant data-driven nature of possibly-gapped histogram.

\subsection{Decoding errors}
Next we consider the decoding error coming from Uniform data-compression. First let the identically independently distributed (i.i.d) random variables be denoted as $\{X_i\}^n_{i=1}$, that is, $X_i$ is distributed with respect to (w.r.t) Standard Uniform density function $f(x)=F'(x)=1, \; \forall x\in [0,1]$. Denote the order statistics as $\{X_{(i)}\}^n_{i=1}$, and the density of $k-$th order statistic $X_{(k)}$, say $g_k(x)$, is evaluated as below.
\begin{eqnarray*}
&&Pr[x< X_{(k)}<x+\delta]\\
&=& C^n_k [F(x)]^{k-1}[F(x+\delta)-F(x)][1-F(x+\delta)]^{n-k}\\
 &\approx&  C^n_k [F(x)]^{k-1}f(x)[1-F(x+\delta)]^{n-k}\delta \\ &=& g_k(x)\delta \\
\text{so that, } g_k(x)&=&C^n_k x^{k-1}[1-x]^{n-k}.
\end{eqnarray*}
Here $g_k(x)$ becomes the exponential density for the extreme ordered statistics $X_{(1)}$ ($k=1$) and $X_{(n)}$ ($k=n$).

Then we have,
\begin{eqnarray*}
E[X_{(k)}] &=& \frac{k}{n+1};\;
E[X^2_{(k)}] = \frac{k(k+1)}{(n+1)(n+2)};\\
Var[X_{(k)}]&=& E[X^2_{(k)}]-\{E[X_{(k)}]\}^2=\frac{k(n-k+1)}{(n+1)^2(n+2)}.
\end{eqnarray*}
Also, we have,
\[
\sum^n_{k=1}Var[X_{(k)}]=\frac{n}{6(n+1)}.
\]

Given the observed data as $\{x^o_{(i)}\}^n_{i=1}$, further let $\tilde{X}_i \sim U[0,1]$ also be another set of observations, distributed with standard Uniform distribution. Then Decoding Error Sum of Squares(DESS) is evaluated as:
\begin{eqnarray*}
\sum^n_{k=1}[\tilde{X}_{(k)}-x^o_{(k)}]^2&=&\sum^n_{k=1}[\tilde{X}_{(k)}-E[\tilde{X}_{(k)}]+E[\tilde{X}_{(k)}]-x^o_{(k)}]^2 \\
&=&\frac{n}{6(n+1)}+\sum^n_{k=1}[x^o_{(k)}-\frac{k}{n+1}]^2 \\
& \approx & \frac{1}{6} +\frac{1}{6} = \frac{1}{3},
\end{eqnarray*}
since we expect that the second term in the last equation is about the same size of $1/6$ if $\{x^o_{(i)}\}^n_{i=1}$ are realized from i.i.d $U[0,1]$.

In general, by a  Uniform-part $U_p[a,b]$, we denote a Uniform distribution on $[a,b]$ in a possibly-gapped histogram. Thus a set of observations in a bin $[a,b]$ in a histogram will be from $U_p[a,b]$, if its decoding-error is about $\frac{1}{3}(b-a)^2$. This is a definite property requirement, which we call the `DESS criterion'. Since this requirement is independent of sample size for large $n$, it must be satisfied by all parts of the possibly-gapped histogram. It is clear that the majority of candidates in the ensemble of Two-layer one dimensional Ising model is not feasible.

The next issue is the fact that an Uniform-part $U_p[a,b]$ can be divided into several Uniform-parts $U_p[a_j,b_j]$ with $a_j=b_{j-1}$ for all $j=1, .., J$. Then the total DESS of the collection $\{U_p[a_j,b_j]\}^J$ is about
\[
1/3\sum^J_{j=1} (b_j-a_j)^2 (<< 1/3(b-a)^2)
\]
which can be much smaller than the DESS $(\approx 1/3(b-a)^2)$ of the original $U_p[a,b]$.

On the other hand, this collection of $\{U_p[a_j,b_j]\}^J$ incurs increasing cost of the coding length for the boundaries $\{[a_j,b_j]\}^J$. So it is necessary to balance between DESS due to stochastic randomness, and complexity of deterministic structure. Let $L_0$ be the relative index between the costs:  decoding error and coding length. Based on the criterion of model simplicity over complexity, we need to make sure that local boundary points $a_j$s or $b_j$s are chosen, if the following inequality holds:
%\begin{eqnarray*}
\[
 1/3(b-a)^2 > 1/3\sum^J (b_j-a_j)^2+ (J-1)L_0;
 \]
 or,
 \[
 1/3\sum^J_{j\neq j'} (b_j-a_j)(b_{j'}-a_{j'})> (J-1)L_0;
 \]
%\end{eqnarray*}
where $L_0$ is an specific index measuring how many units of decoding error is ``equal to'' the coding length cost of one boundary point. That is, $L_0$ is determined within the domain system and is independent of sample size $n$. Thus this balancing criterion being independent of $n$ is realistic and practical, but is at odds with statistical modeling and its model selection in the Statistics literature. For instance, the principle of Minimum Description Length(MDL) is to minimize with respect to $k$:
\[
MDL(k)=-log(P_k(x|\hat{\theta}))\pi(\hat{\theta})+\frac{k}{2}\log{n} + {O}(k),
\]
where both terms on the right-hand side grow with sample size $n$. The first term accounts for the predictive errors, while the second term accounts for the $O(\frac{1}{\sqrt{n}})$ precision of all global parameter estimators. Here we would like to point out a fundamental assumption: {\bf the ensemble of candidate models is independent of sample size $n$}. This assumption is always imposed implicitly, but hardly spelled out  explicitly in the statistical literature. One serious implication of this assumption is the homogeneity of data structure that does not change as sample size increases. This homogeneity is apparently and practically not possible to hold in histogram construction.

\subsection{Possibly-gapped histogram and 1st phase of ANOHT}
Given a value of index $L_0$,  an observed data $\{x^o_{(i)}\}^n_{i=1}$, and the fact that the exhaustive search for an optimal possibly-gapped histogram within the ensemble prescribed by the Two-Layer one dimensional Ising model is overwhelmingly impossible, one practically feasible computational solution is to apply the Hierarchical Clustering (HC) algorithm with the recommended Complete module. This choice of the module provides the needed subdividing tendency. This tendency fits the Uniform distribution well in the sense that subdividing Uniform distribution reduces the total decoding errors.

The computational algorithm for a small set of potentially possibly-gapped histograms is proposed based on the bifurcating feature of an HC-tree as follows. Let's define a bifurcating inter-node in an HC-tree to be active, if  its parent inter-node has not been marked with a STOP sign. For practical purpose, the index $L_0$ is used as a threshold value.\\
%\subsection*{}
%\iffalse
 \begin{algorithm}
 \caption{Algorithm for possibly-gapped histograms}
\begin{algorithmic}
\State \textbf{{Step-1:}} Compute the DESS on the single data-range, $[x^o_{(1)},x^o_{(n)}]$, as having only one bin. If DESS-criterion is satisfied, then stop this algorithm. If DESS criterion is violated, go to the Step-2.

\State \textbf{{Step-2:}}  Find the next highest active bifurcating inter-node and compute its DESS, say $DESS(P)$. If $DESS(P) < L_0$, or  DESS criterion is satisfied, then mark this inter-node with a STOP sign. Then check if there are active inter-nodes remaining in the HC-tree. If yes, then repeat to Step-2, otherwise go the Step-3. If $DESS(P) \geq L_0$, and if DESS criterion is violated, then repeat Step-2.

\State \textbf{{Step-3:}} Stop when no more active inter-nodes are left. Check the existential gap between all consecutive Uniform-parts.
\end{algorithmic}
\end{algorithm}
%\fi
There are several ways to check whether a gap exists between two consecutive Uniform parts of a histogram. For example, theoretically extended boundaries can be estimated via Exponential distribution of the extreme random variable. If the two estimated boundaries don't cross each other, then the two Uniform parts are separated. So there exists a gap between them. Another practical way of checking is to recalculate the DESS of these two Uniform parts after modifying both of them so that they share a common extended boundary, i,e. the mid-point of their extreme values. Then if both of them satisfy the DESS criterion, then there doesn't exist a gap. Otherwise, there exists one.

The histogram resulted from the above algorithm is the coarsest version of possibly-gapped histogram. If an optimal histogram is needed, then all Uniform parts with large DESS $(> L_0)$ can be bifurcated according to HC-tree branches to further improve the total decoding errors at a cost $L_0$ of coding one more boundary.

Further, the higher an HC-tree inter-node is, the bigger is the gap potential. In fact, a gap is often visible in the coarsest version of a possibly-gapped histogram, since it gives rise to one possibly-gapped piece-wise  linear approximation onto the empirical distribution function, see illustrations of four features of Iris data in Fig. \ref{fig:hist_iris}. Particularly in the third and fourth panels of Fig. \ref{fig:hist_iris} for Petal Length and Petal width, we see evident gaps. The biological significance of the possibly-gapped histograms of Iris' Petal length and width is clearly revealed through the separation of color-coded species.

Here we explicitly illustrate such computations on Iris data set, after standardizing each feature to zero mean and unit standard deviation. When a histogram is indeed ``gapped", this gap should be identified as a corresponding break in the empirical CDF also. We can verify this through the distributions of the extreme random variables of the $U(a,b)$ distribution, as the following.
\[
\hat{a}=X_{(1)}-\frac{(X_{(n^*)}-X_{(1)})}{n^*+1} \text{ \hspace{0.3 cm} and \hspace{0.3 cm} } \hat{b}=X_{(n^*)} + \frac{(X_{(n^*)}-X_{(1)})}{n^*+1},
\]
$n^*$ being the bin-frequency.

As in Fig. \ref{fig:hist_iris}(c), the right boundary of the first bin of petal length can be estimated as $-1.032906$, while the left boundary of the second bin is estimated as $-0.5127225$.
Also, in Fig. \ref{fig:hist_iris}(d), the right boundary of the second bin and the left boundary of the third bin of petal width are estimated as $-0.7274579$ and $-0.3240107$. The computed left and right boundary estimates do not cross each other in both cases. Hence this implies that the gap between Setosa  and the other two species is significant. The DESS values of the final gapped bins are pretty low and fall around the uniform DESS criterion. In contrast, the Iris' Sepal length and width do not bear such significance.

As a histogram of a feature is usually taken as a simple data-visualization step within a long and complicated process of data analysis, it hardly constitutes any stand-alone goal of data analysis. Here we would like to point out that this impression is completely incorrect. In fact, a possibly-gapped histogram indeed is an important and useful tool for data analysis.

Consider a possibly-gapped histogram constructed via Algorithm 1 applied to a data set by pooling measurements from $J$ treatments. We are ready to simultaneously compare these $J$ treatment-specific distributions by encoding $J$ treatment-specific colors onto each bin with respect to its member-measurements' treatment identifications.  Each bin has a composition of colored proportions. So its entropy relative to the entropy of $J$ treatment sample sizes, or their ratio is an index for bin-specific local comparison among the $J$ treatments. A p-value can be also derived from the simulation study via simple random sampling without replacement. Hence we can afford to test these $J$ treatments by pointing out where and how they are different. If an overall index is needed, then the weighted entropy across all bins is calculated, so is its p-value.  This is the first phase of ANOHT.

In summary, a color-coded possibly-gapped histogram can simultaneously offer more informative comparisons among $J$ treatments than Kolmogorov-Smirnov test, which is limited for pairwise distributional comparisons, and One-Way Analysis of Variance (ANOVA), which focus on comparing the $J$ mean-values.

This first phase of ANOHT on Iris data is shown through the middle column panels of Fig. \ref{fig:hist_iris}.  All four color-coded possibly-gapped histograms reveal very informative comparisons among the three Iris species. The compositions of bin-specific color-codes very importantly reveal how and where these species are different. Such locality information is essential and invaluable. The biological meaning of gaps in histograms pertaining to Petal length and width become evident and crucial. Here, for local testing purpose with respect to all four features, we see that majorities of p-values are either zeros or extremely small due to single color dominance or one color being completely missing. Likewise, the overall p-values are extremely small. Therefore, biologically speaking, these four features all provide informative pieces of information for differentiating these three species of Iris. This is a conclusion that has not been seen in literature, particularly in machine learning.

\subsection{Tree geometry and 2nd phase of ANOHT}
As the first phase of ANOHT provides a way of simultaneous comparison of $J$ treatments locally and globally, the second phase of ANOHT intends to precisely address issues regarding: which treatments are closer to which and farther away from which. Our algorithm will construct a tree-geometry upon these $J$ treatment-nodes to provide comprehensive information regarding these issues. Upon this tree geometry, this algorithm further evaluates an ``authenticity'' index at every inter-node of tree-branch. Here an authentic tree branch means that its memberships are strongly supported by the data, not due to chance.

Heuristic ideas underlying this algorithm are as follows. A color-coded possibly-gapped histogram (of counts) is transformed into a matrix: one colored bin is turned into one column according to color-specific rows, that is, one color for one row. Then the $J \times K$ matrix of counts is normalized row-by-row into a matrix having $K$ columns of frequencies. Then, for simplicity, Euclidean distance is calculated between any pair of rows and the  Hierarchical clustering algorithm is applied to build an HC-tree upon the row axis. So a heatmap superimposed with such an HC-tree has resulted with $J-1$ inter-nodes, so $J-1$ branches.

Along the construction process of this HC-tree, each of $J-1$ inter-node is associated with a tree height. We sort and digitally rank these $J-1$ tree heights from smallest to largest as a way of re-normalization. Hence each inter-node specifying a branch will be assigned with a rank-digit.  This inter-node-specific rank-digit indicates that the tree height for any node within this branch to meet with any other node outside of this branch must be ranked higher.

Using the Empirical Process Theory for complete data (\cite{Fushing95}, \cite{Shorack}), mimicking is then applied on this heatmap, by simulating each row vector. The specific multivariate Normality structure used here is given as follows . Let $F_n(t)=1/n\sum^n_{i=1} \mathbbm{1}(X_i \leq t)$ be a generic empirical distribution estimating a true distribution $F(t)$ pertaining to one of the $J$ treatments. The classical empirical process theory implies that
\[
\sqrt{n}(F_n(t)-F(t)) \xrightarrow[]{n \rightarrow \infty} Z_F(t), \forall t
\]
where $Z_F(t)$ is a Brownian Bridge, which is a Gaussian process with covariance function $\sigma_F(t,t')=F(t)(1-F(t'))$ when $t<t'$. Therefore, with $(t_0, t_1, t_2, ...., t_K)$ being the chosen ordered boundaries of $K$ bins, the vector $(Z(t_1),\hdots Z(t_K))^T$ comes from a family of Multivariate Normal distribution as below.

\[
\tilde{Z}(t) = (Z(t_1),\hdots Z(t_K))^T \sim N(0,\Sigma_K) \text{  with  } t_1<t_2< \hdots <t_K
\]

where,

\begin{eqnarray}
\label{eqn:*}
\Sigma_k =&
\left(
\begin{array}{cccc}
F(t_1)(1-F(t_1)) & F(t_1)(1-F(t_2)) & \hdots & F(t_1)(1-F(t_K)) \\
& F(t_2)(1-F(t_2)) & \hdots & F(t_2)(1-F(t_K)) \\
&  & \hdots & \hdots \\
&  &  & F(t_K)(1-F(t_K))
\end{array}
\right)   \nonumber \\
=& A \text{ diag}(F(t_1) , F(t_2) , \hdots , F(t_K)) A^T -\tilde{F}(t)\tilde{F}^T(t)
\end{eqnarray}
\iffalse
$
\iffalse
=&
\left(
\begin{array}{cccc}
1 & 0 & \hdots & 0 \\
1 & 1 & \hdots & 0 \\
\vdots & \vdots & \ddots & \vdots \\
1 & 1 & \hdots & 1
\end{array}
\right)
\left(
\begin{array}{cccc}
\Delta_1F(t) & 0 & \hdots & 0 \\
0 & \Delta_2F(t) & \hdots & 0 \\
\vdots & \vdots & \ddots & \vdots \\
0 & 0 & \hdots & \Delta_kF(t)
\end{array}
\right)
\left(
\begin{array}{cccc}
1 & 1 & \hdots & 1 \\
0 & 1 & \hdots & 1 \\
\vdots & \vdots & \ddots & \vdots \\
0 & 0 & \hdots & 1
\end{array}
\right)   \nonumber \\
&-
\left(
\begin{array}{c}
F(t_1) \\
F(t_2)\\
\vdots \\
F(t_K)
\end{array}
\right)
\left(
\begin{array}{cccc}
F(t_1) & F(t_2) & \hdots & F(t_K)
\end{array}
\right)  \nonumber \\
\fi$
\fi

\noindent with $\Delta_k F(t) = F(t_k)-F(t_{k-1}), k=1,\hdots, K.$ $\tilde{F}^T(t) = (F(t_1) , F(t_2) , \hdots , F(t_K))$, and  
the matrix $A$ and its inverse $A^{-1}$ taking the following forms,

$A=
\left(
\begin{array}{cccc}
1 & 0 & \hdots & 0 \\
1 & 1 & \hdots & 0 \\
\vdots & \vdots & \ddots & \vdots \\
1 & 1 & \hdots & 1
\end{array}
\right) \text{ and }
A^{-1} =
\left(
\begin{array}{cccccc}
1 & 0 & 0 & \hdots & 0 & 0 \\
-1 & 1 & 0 & \hdots & 0 & 0 \\
0 & -1 & 1 & \hdots & 0 & 0 \\
\vdots & \vdots & \vdots & \vdots & \vdots  & \vdots \\
0 & 0 & 0 & \hdots & 1 & 0 \\
0 & 0 & 0 & \hdots & -1  & 1
\end{array}
\right)
$

\noindent Thus we have $A^{-1}\tilde{Z}(t) = (\Delta_1Z(t), \Delta_2Z(t), \hdots, \Delta_K Z(t) )^T = \Delta \tilde{Z}(t) $.

\noindent Further, we have the following asymptotic multivariate Normality result based on equation \eqref{eqn:*}

\[
\Delta \tilde{Z}(t) \sim N(0, \Sigma^*_K),
\]

with
\begin{eqnarray}
\Sigma^*_K =& diag(\Delta_1F(t), \Delta_2F(t), \hdots, \Delta_K F(t) ) - \Delta\tilde{F}(t) \left( \Delta\tilde{F}(t)\right)^T \nonumber
\end{eqnarray}

\iffalse
$
=& \left(
\begin{array}{cccc}
	\Delta_1F(t)(1-\Delta_1F(t)) & -\Delta_1F(t)\Delta_2F(t) & \hdots & -\Delta_1F(t)\Delta_K F(t) \\
	\ast & \Delta_2F(t)(1-\Delta_2F(t)) & \hdots & -\Delta_2F(t)\Delta_K F(t) \\
	\ast &  \ast & \ddots & \vdots \\
	\ast & \ast & \ast & \Delta_K F(t)(1-\Delta_K F(t))
\end{array}
\right) $
\fi
\noindent This $K\times K$ covariance matrix $\Sigma^*_K$ with small and negative off-diagonal entries: $-\Delta_j F(t)\Delta_{j'} F(t),\; j\neq j'$, is the key component for mimicking each row vector of the aforementioned $J\times K$ matrix (or heatmap). An HC-tree is also derived for each mimicked heatmap. By performing such mimicking procedure many times, an ensemble of HC-trees is generated. With this ensemble, we are able to compute an authenticity index at each inter-node on the original HC-tree by counting the proportion of mimicked HC-tree having the following property: the rank-digit pertaining to the smallest branch containing all nodes belonging to the original branch being smaller or equal to the original rank-digit defining the original branch. See also the description of Algorithm 2 below.

%\iffalse
\begin{algorithm}
\caption{ANOHT}
\begin{algorithmic}
\State \textbf{{Step-1:}} Build a possibly gapped histogram on the pooled data, using Algorithm 1.

\State \textbf{{Step-2:}} Compute a matrix with the frequency of each species in the bins constructed in the gapped histogram of pooled data. For $j=1,...J$ treatments and $k=1,...,K$ bins $T_{jk}$ denotes the number of subjects of $j$-th treatment falling in $k$th bin. $n_j=\sum_{k=1}^K T_{jk}$= total frequency of $j$-th treatment. Define $P_{jk}=\frac{T_{jk}}{n_j}$, $\forall j, k$. Build a tree on the row-axis of the $J\times K$ matrix $T$. This tree will be the point of reference.

\State \textbf{{Step-3:}} Mimicking $T$ by randomly simulating a $J\times K$ matrix $M=[M_{jk}]$ with its $j$-th row $\widetilde{M}[j,.]$ being generated with respect to
\[
M[j, .]=(n_j \hat{P}_{j1}, ... , n_j \hat{P}_{jK})=n_j\widetilde{\hat{P}}_j, \text{where }  \widetilde{\hat{P}}_{j} \sim \text{Multi-Normal} \left(\widetilde{P}_{j}, \frac{1}{n_j}\Sigma^*_K \right)
\]

\State \textbf{{Step-4:}} Generate the matrix $M=[M_{jk}]$ 10,000 times and for each generated matrix build a tree on the rows and check if the branches of these new trees are similar to the branching of the reference tree. For each bifurcating inter-node of the reference tree, calculate an authenticity index as the percentage of times the nodes re-appeared together in the mimicked trees.
\end{algorithmic}
\end{algorithm}
%\fi
It is emphasized once more that the algorithmically computed tree geometry would allow us to see which branch of treatment-nodes is authentic in the sense that within-branch distances are smaller than between-branch distances with significantly high probabilities. In other words, the formation of branch-specific memberships is not likely due to noises. This authenticity index evaluates and confirms potential biological or mechanistic basis for this branch. This serves as one part of the knowledge discovery.

\begin{figure}
\begin{center}
\includegraphics[height=6in, width=6in]{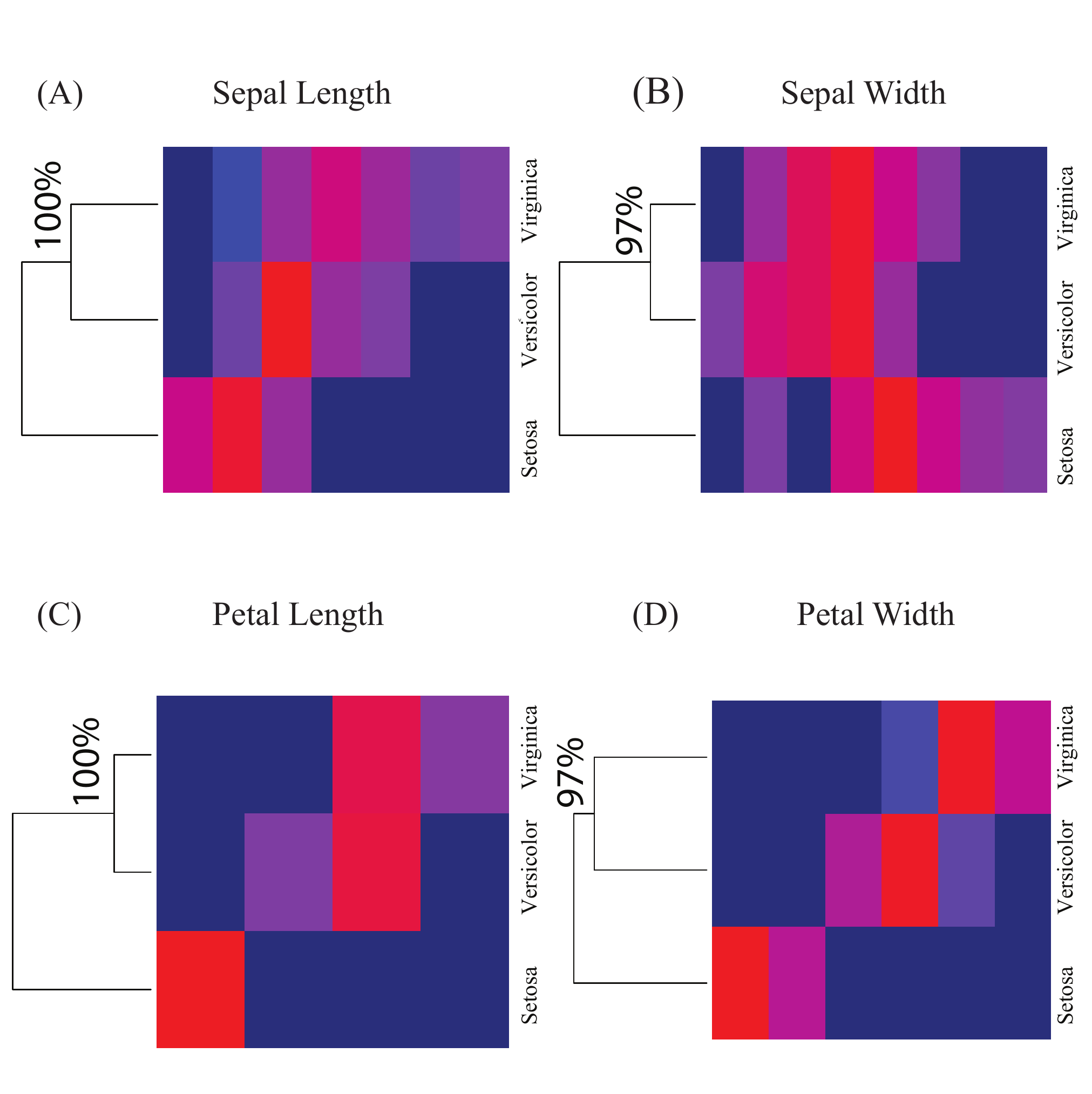}
\end{center}
\caption{ANOHT for all columns of iris data set. Out of $10,000$ re-generations, the nodes Versicolor and Virginica stay together $100\%$ of the times  in the Sepal length and Petal length trees($97\%$ for Sepal and Petal width trees), and in those cases Setosa stays in a different branch.  A color closer to red represents higher value and dark blue corresponds to the lowest frequency zero.}
\label{fig:ANOHT_iris}
\end{figure}

By individually applying the Algorithm 2 based on all four histograms in Fig. 2, we obtain four tree geometries on three Iris species-nodes with respect to their four features, as shown in Fig.\ref{fig:ANOHT_iris}. In
each of the four features of Iris, about $97\%$ to $100\%$ of the 10,000 generations of the species tree confirm that the branch of Virginica and Versicolor  stayed together, and the Setosa stayed separated from them. Further, the heatmap clearly pinpoints where are the significant differences, so that the species Setosa is rather distinct from the two other species. It is not unreasonable to think that each of these four HC-trees with authenticity indexes will shed light on the phylogenetics of these three species. This is one of the most  significant merits of the 2nd phase of ANOHT.

\section{ANOHT on right censored data.}
When data is compromised, such as being right-censored in survival analysis, the construction of a histogram is rarely seen in practice (\cite{Kleinbaum}) as well in literature (\cite{Cox}, \cite{Andersen}, \cite{Kalbfleisch}). We propose to construct a possibly-gapped histogram and perform ANOHT based on Kaplan-Meier estimate of a survival function. Then the ANOHT is also performed based on the Nelson-Allen estimate of the cumulative hazard function. The later analysis is much simpler and easier to apply than the former analysis. We apply both analyses on a heavily censored divorce data.

The foundation for the 1st phase of ANOHT is the fact that the Kaplan-Meier estimate has weights only on uncensored time points, see \cite{Miller}. And the weighting scheme is termed redistribution-to-the-right, see \cite{Efron}. Therefore the first phase of ANOHT can be carried out by building a possibly-gapped histogram on uncensored time points, and then applying the redistribution-to-the-right weighting scheme to adjust the weights on all bins. Nonetheless, the 2nd phase of ANOHT would need a bit more complicated empirical process theoretical results, which are given below.

\subsection{Empirical process theory on Kaplan-Meier estimate of the survival function.}
Given a treatment of interest, let the right-censored data set be generically denoted as $\mathcal{X} \{ (X_i,\delta_i) \}_{i=1}^n$ with $X_i = T_i\wedge C_i, \delta_i = \mathbbm{1}_{\{ T_i \leq C_i \}} $ and $T_i \perp C_i $(independent) for all $i = 1,\hdots, n.$ The Kaplan-Meier estimate of the survival function $S(t) = P(T>t)$ is constructed as

\[
\hat{S}_n(t) = \prod\limits_{X_{(i)} \leq t} \left( 1-\frac{\delta_{(i)}}{n-i+1} \right)
\]

\noindent where $\{X_{(i)} \}_{i=1}^n$ is the order-statistics ranking from the smallest to the largest, and $\delta_{(i)}$ is the corresponding censoring status of $X_{(i)}$.

\noindent The right censored version of empirical process theory (\cite{Miller}, \cite{Andersen},\cite{Kalbfleisch}) states that
\[
\sqrt{n}(\hat{S}_n(t)-S(t)) \xrightarrow[]{n \rightarrow \infty} Z_s(t), \forall t \in \mathbb{R}^+
\]
where $Z_s(t)$ is a Gaussian process with characteristic covariance function
\[
\sigma_s(t_1,t_2) = S(t_1)S(t_2) \int_{0}^{t_1 \wedge t_2} \frac{dF_u(s)}{(1-H(s))^2}
\]
with $F_u(t)= P(T_i \leq t,\delta_i = 1)$ and $1-H(t) = P(X_i >t) = P(T_i>t, C_i>t)$. It is noted that $\sigma_F(t_1,t_2)  = \sigma_s(t_1,t_2)$ when there is no censoring.

Likewise the $K$-dimensional vector $\tilde{Z}_s(t) = (Z_s(t_1),\hdots, Z_s(t_K))^T$ is asymptotically normally distributed, that is:
\[
\tilde{Z}_s(t) \sim N(0, \Sigma^{\#}_K)
\]
where
\begin{eqnarray}
\Sigma^{\#}_K &=& diag(S(t_1),\hdots, S(t_K))\cdot A \cdot diag\left( \int_0^{t_1} \frac{dF_u(s)}{(1-H(s))^2}, \int_{t_1}^{t_2} \frac{dF_u(s)}{(1-H(s))^2}, \hdots  \right)\nonumber \\
&\cdot&A^T \cdot diag(S(t_1),\hdots, S(t_K)). \nonumber
\end{eqnarray}
Hence we have the asymptotic normality of the k-dim vector $ \Delta\tilde{ Z}(t) \sim N(0, A^{-1}\Sigma^{\#}_k (A^{-1}))$. Throughout this section we employee the following approximation:
\[
\int_{t_{j-1}}^{t_j} \frac{dF_u(s)}{(1-H(s))^2} \approx n \sum\limits_{t_{j-1}\leq X_{(i)} \leq t_j} \frac{\delta_{(i)}}{(n-i)(n-i+1)}.
\]

It is clear that the Kaplan-Meier estimate $\hat{S}_n(t)$ has jumps only on uncensored time points. Hence a construction of a possibly-gapped histogram needs only to take one more step of readjusting the weighting, that is, a histogram for a right-censored data set can be constructed via the following two steps
\begin{enumerate}
	\item Build a histogram upon uncensored(or complete) data points;
	\item Re-adjust the weights by applying the redistribution-to-the-right scheme for constructing the Kaplan-Meier estimate.
\end{enumerate}

\subsection{Empirical process theory on Nelson-Aalen estimate of cumulative hazard function.}
Due to the Martingale central limit theory, see \cite{Andersen} and \cite{Kalbfleisch}, the Nelson-Aalen estimate of the cumulative hazard function $\Lambda(t) = -\ln S(t)$ is popularly used in survival analysis as an alternative to Kaplan-Meier estimate in statistical inferences, see \cite{Fushing96a}, \cite{Fushing96b} and \cite{Fushing01}.

The Nelson-Aalen estimate of $\Lambda(t)$ is denoted as:
\[
\hat{\Lambda}_n(t) = \sum\limits_{X_{(i)} \leq t} \frac{\delta_{(i)}}{n-i+1}, \forall t \in \mathbb{R}^+
\]
and the Martingale Central Limit Theorem assures that
\[\sqrt{n}(\hat{\Lambda}_n(t)-\Lambda(t) ) \xrightarrow[]{n \rightarrow \infty} Z_\Lambda(t)
\]
where $Z_\Lambda(t)$ is a Gaussian process with independent increment property. That is, its covariance function is
\[
\sigma_\Lambda(t_1,t_2) = \int_{0}^{t_1 \wedge t_2} \frac{dF_u(s)}{(1-H(s))^2} = \sigma(t_1 \wedge t_2)
\].

Hence $Z_\Lambda(t)$ acts like a Brownian motion, so the components of the k-dimensional vector $\Delta \tilde{Z}_\Lambda(t) = (\Delta_1Z_\Lambda(t), \hdots, \Delta_K Z_\Lambda(t))$
are indeed independently distributed, that is,
\[
\Delta\tilde{Z}_\Lambda(t) \sim B(0, \Sigma_K^{**})
\]
with
\[
\Sigma_K^{**} = diag \left(\int_{t_0}^{t_1} \frac{dF_u(s)}{(1-H(s))^2}, \int_{t_1}^{t_2} \frac{dF_u(s)}{(1-H(s))^2}, \hdots \int_{t_{K-1}}^{t_K} \frac{dF_u(s)}{(1-H(s))^2}\right).
\]

Here $\Sigma^*_K$ is the foundation for the 2nd phase of ANOHT on complete data, while $\Sigma^{\#}_K$ and $\Sigma_K^{**}$ are the foundations for the 2nd phase of ANOHT on right censored data.

\section{Results on baseball data}

In this section, we apply our algorithms to construct possibly-gapped histograms and perform Analysis of Histogram (ANOHT) on pitching data of three well-known pitchers in Major League Baseball (MLB). The three pitchers are Jake J. Arrieta (Chicago Cubs), Kyle C. Hendricks (Chicago Cubs) and Robert A. Dickey (Atlanta Braves). The first two pitchers with rather distinct pitching styles were keys to the 2016 World Series Champion won by Chicago Cubs, whose previous titles were in 1908 and 1907.  The first pitcher is the 2015 Cy Young Award winner, the second one, who graduated from Dartmouth College, has a nickname ``The Professor", while the third one is the first knuckleball pitcher to win the Cy Young Award. This pitching data set contains 18732 pitches in the 2015-2016 season, including Arrieta's 6848 pitches.

The data set was downloaded from in the MLB official website provided by PITCHf/x system. This system is installed in all 30 MLB stadiums to track and digitally record the full trajectory of live baseball pitch since 2006. The data contains 21 features of each and every single pitch and batting results throughout every single game in the regular season. Here we only look at two important pitching characteristic features: start-speed and break-length.  The start-speed is the detected speed of a baseball at the release point of a pitch, while break-length is the measured largest distance from the baseball's curved trajectory to the straight-line linking its release point to the front of home plate. These two features are main characteristics of a pitch. They are related to each other in rather distinct ways among 8 computer classified pitch-types: 1) Four-seam Fastball (FF); 2) Fastball cutter (FC); 3) Fastball sinker(SI); 4) Slider (SL); 5) Changeup (CH); 6) Curveball (CU); 7) Knuckleball (KN); 8) Eephus (EP). The first three pitching types are in the category of fastball, while the last five are in the category of off-speed pitches.  It needs to be kept in mind that a pitcher's repertoire of pitch-types is only a strict sub-set of these eight pitch-types.

The start-speed of Arrieta's fastball can go up to nearly 100 mph (miles per hour), while Dickey's knuckleball can be as slow as 65mph. It is known in general that a pitch with higher start-speed tends to have smaller break-length. In fact, their relationship is relatively nonlinear. Hence, combinations of these two features are usually cleverly crafted by every professional baseball pitcher in order to effectively face batters. So it is of great interest to compare one single pitcher's as well as multiple pitchers' distinct pitch-types from these two aspects.

\begin{figure}[hbtp]
\begin{center}
	\includegraphics[scale=0.5]{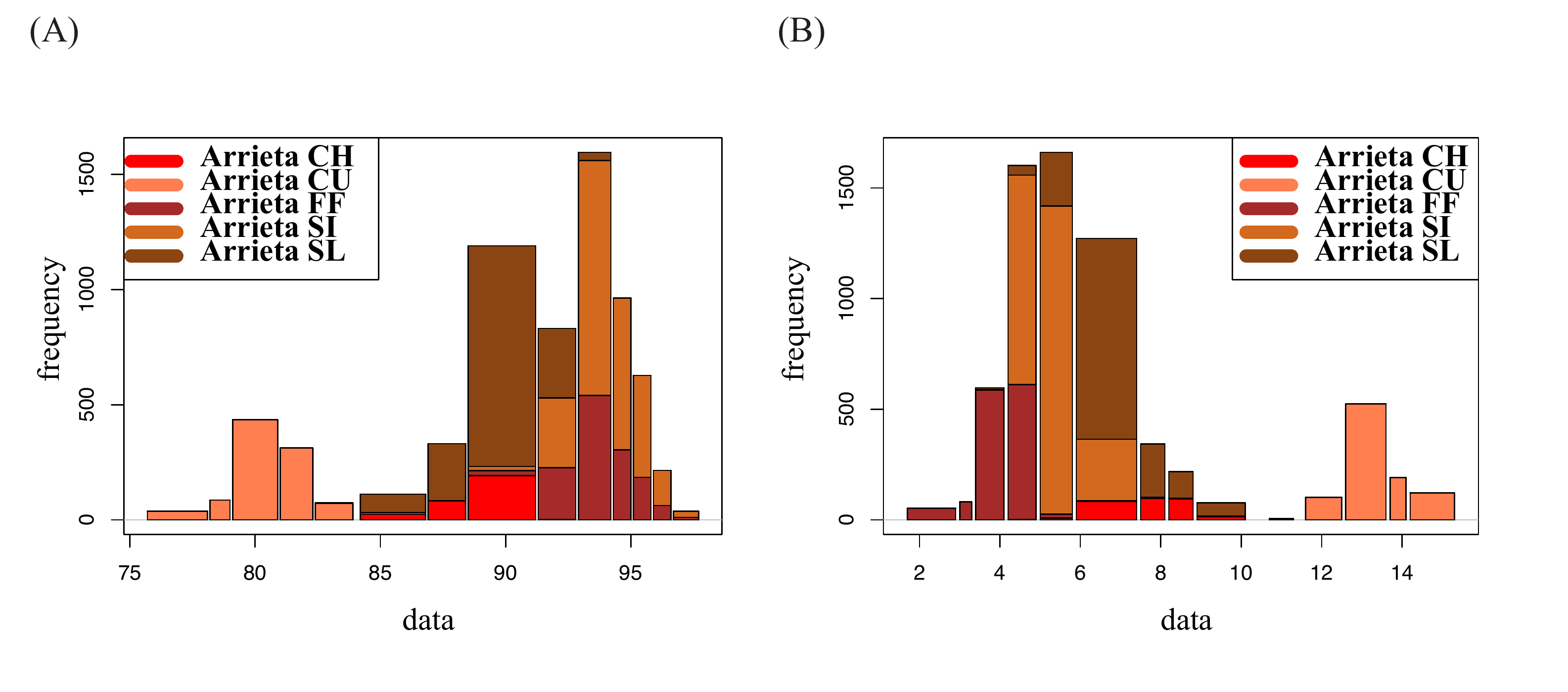}
\end{center}
	\caption{Possibly-gapped histograms for the start-speed(A) and break-length(B) measurements of pitches thrown by Jake Arrieta in 2015-2016. The total number of pitches from 3 pitchers is 18732 and out of them 6848 are from Arrieta. The five different colors in the histograms show the frequency of five different pitching styles of Arrieta in each bin. }
	\label{fig:hist_arrieta}
\end{figure}

\subsection{Data analysis on one single pitcher's pitch-types.}
We begin with our first phase of ANOHT on comparing pitch-types of a single pitcher from the start-speed and break-length aspects.  For instance, to compare Arrieta's 5 pitch-types: {FF, SI, SL, CH, CU}, we apply the Algorithm 1 to construct two possibly-gapped histograms of start-speed and break-length, as shown in the panels (A) and (B) of Fig.\ref{fig:hist_arrieta}, respectively. In panel (A), we see a clear gap around 84mph. This gap bears a mechanistic difference between his curveball (CU) and the rest of 4 pitch-types, which have significant higher speed. That is, this pitch-type is purposely and distinctively carried out by this pitcher.  Hence this gap indicates a clear-cut implementation due to the pitcher's control capability.

This capability is even more strikingly demonstrated via the gap shown in the histogram of break-length, as shown in panel (B). Again the pitch type CU is completely separated from the rest of 4 types. It is also evident that pitch-types FF and SI are dominant in bins on right extreme of start-speed histogram and correspondingly dominant in bins on the left-extreme of break-length histogram.

In summary these two color coded possibly-gapped histograms clearly demonstrate the distinctive differences among Arrieta's five pitch-types from the two aspects. In other words, all bin-specific entropies are to be zeros or significantly small in comparison with the entropy of the distribution pertaining to the five categories of pitch-types. That is, all p-values computed via the scheme of simple-random-sampling without replacement are zeros or extremely small. Similar data analysis on the other two pitcher can be likewise carried out.

\subsection{Data analysis on three pitchers' pitch-types.}
\begin{figure}[hbtp]
\includegraphics[height=6in, width=7in]{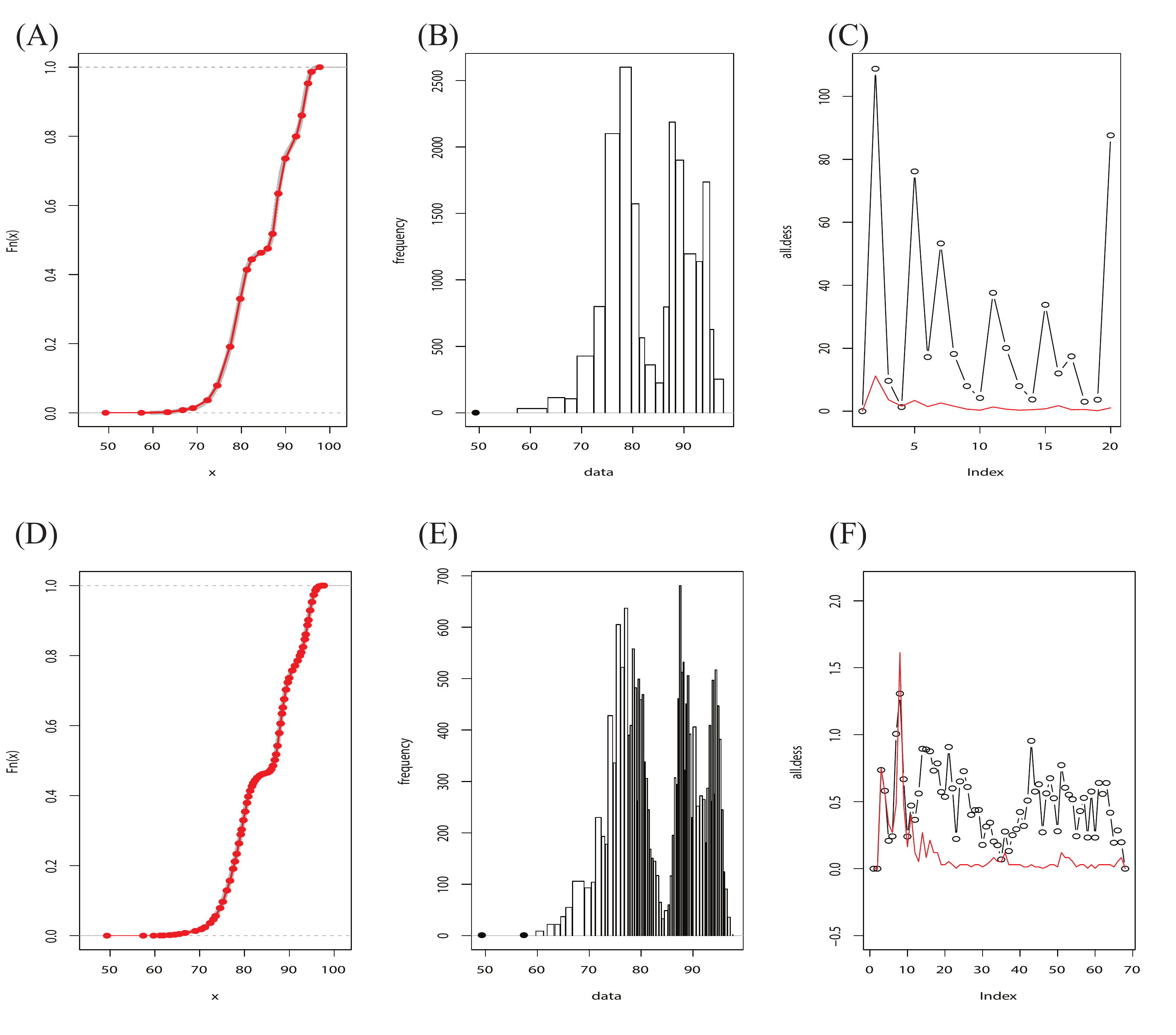}
\caption{Possibly gapped histograms on start-speed measurements of all pitches thrown by Kyle Hendricks, Jake Arrieta and R.A. Dickey in 2015-2016. Sub-figure (A-C) shows a coarse version with high decoding error allowed, and sub-figure (D-F) shows a finer version where decoding error is constrained to be less than 2.}
\label{fig:hist_3pitcher}
\end{figure}

We begin with the first phase of ANOHT to understand how similar or distinct are  pitcher-specific pitch-types from the  aspects of start-speed and break-length.  On top of Arrieta's 5 pitch types, Hendricks has 5 pitch types: {FF, FC, SI, CH, CU}, and Dickey has 3 types: {FF, KN, EP}. So there are 13 pitcher-specific pitch-types (or treatments) in total for comparison. One coarse and one fine resolution of possibly-gapped histograms of start-speed are constructed and reported in two rows of triplet panels of Fig.\ref{fig:hist_3pitcher}, respectively. After applying Algorithm 1 with a choice of $L_0=0.1 \times$tree height, there are 20 bins selected in this coarse version, as seen in panels (A-C).  The piecewise linear approximation on the empirical distribution seems reasonable, as seen in the panel (A), and histogram reveals evident two, or potentially three modes in panel (B). But the all DESS values are relatively high in panel (C).

%[Fig.5 Here]

Therefore, in order to drive DESS values lower, we select a lower $L_0$ and get a histogram with a fine resolution, as shown in panels (D-F). Though the piecewise linear approximation becomes somehow overwhelmingly detailed with 80 bins in panel (D), the histogram is seen with evident 3 modes in panel (E) and corresponding DESS values become significantly smaller than that in panel (C).  That is, the fine resolution histogram with very smaller bins on its right seems to give rise to more detailed distributional structures than the coarse one. This coarse-vs-fine resolution of histograms illustrate why we need to have data-driven bins with data-driven sizes.

\begin{figure}[hbtp]
\begin{center}
	\includegraphics[scale=0.55]{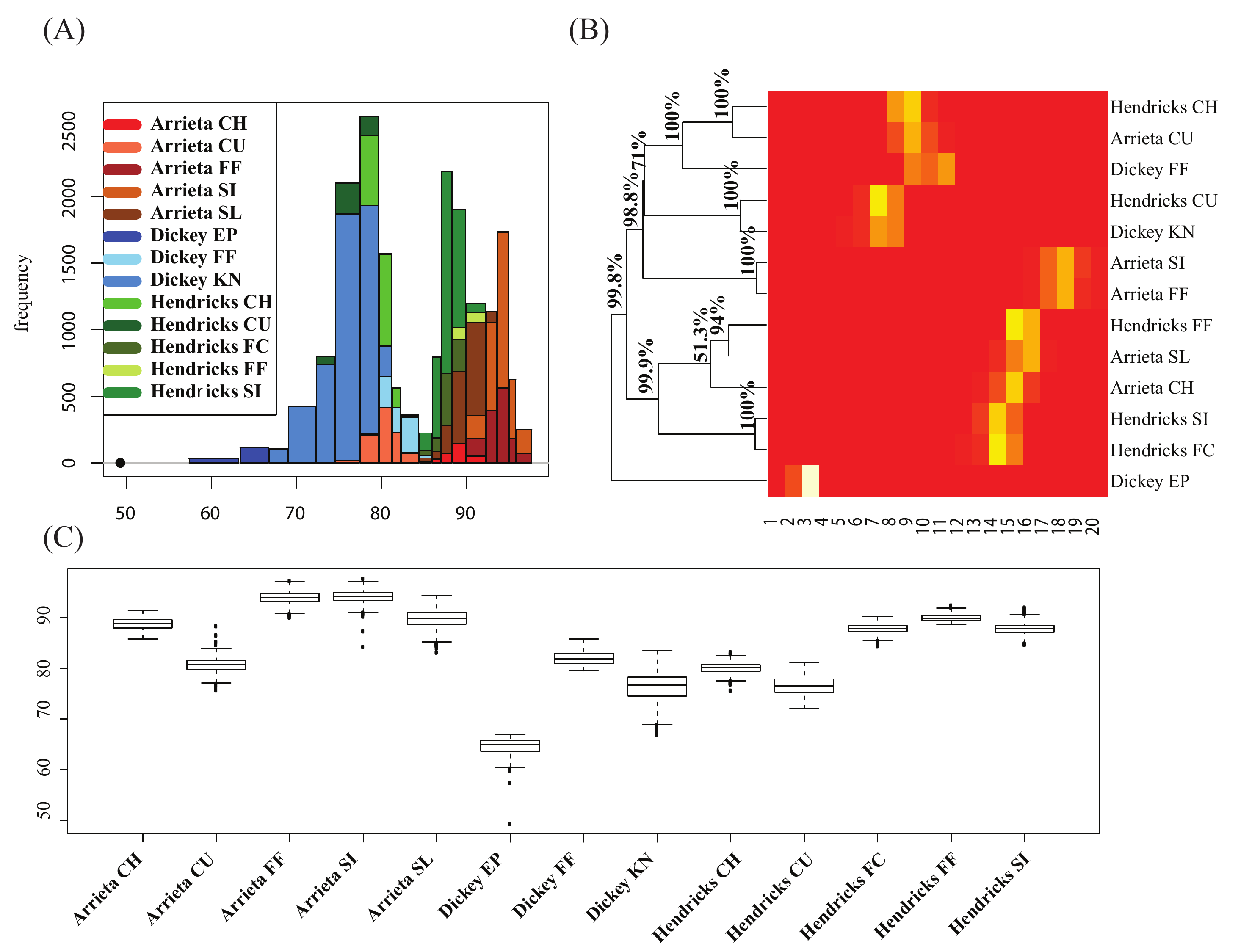}
\end{center}
	\caption{Coarse version of the possibly gapped histogram of start-speed from figure  \ref{fig:hist_3pitcher}, color coded according to the 13 different pitches thrown by the three pitchers in sub-figure (A).
		For differentiating between the 3 pitchers, we used shades of red, blue and green for Arrieta, Dickey and Hendricks respectively . Sub-figure (B) shows the tree on ANOHT matrix with the branch strength percentages, and sub-figure (C) shows the traditional boxplot of all start-speeds of 13 pitching styles.}
	\label{fig:startspeed_ANOHT}
\end{figure}

Then we color code the coarse version of histogram, instead of the fine resolution version, for better visualization when we set to compare these 13 pitcher-specific pitch types, as shown in the panel (A) of Fig.\ref{fig:startspeed_ANOHT}. Upon this histogram of start-speed, we see that bins on its left hand side are dominated by Dickey's KN, and bins on its right hand side are primarily dominated by Arrieta's FF and SI. While Hendrick's pitch types are prevalent on bins situated at both sides of major valley of this histogram. It is clear that, by focusing on where differences actually occur, this color coded histogram is much more informative than traditional boxplot, as shown in panel (C) of Fig.\ref{fig:startspeed_ANOHT}, or ANOVA, which focuses merely on comparisons of mean values.

%[Fig.6 Here]

Then we turn to the second phase of ANOHT to see what the tree geometries of the 13 pitcher-specific pitch-types look like from the aspects of start-speed and break-length. Here it is worth re-emphasizing the significance of tree geometry. Tree geometry is a data-driven structure that makes  performing branch-based-group comparison possible. A branch-based-group specified by an inter-node is confirmed as being authentic, if its branch formation is significantly supported by the data, and not by chance. This confirmation equivalently implies that, with significantly high probability, no tree nodes outside of such a group have smaller distances (or higher similarity) than distances (or similarity) among its group-member-nodes.  From mechanistic perspective, such a concept of authenticity of tree branch (-based-group) has a direct implication on ``phylogenetic'' information contents regarding  the 13 pitcher-specific pitch-types, as would be seen below. From inference perspective, its functions have gone beyond multiple comparisons, such as Tukey's pairwise comparison and others, typically performed in Analysis of Variance (ANOVA) in classical Statistics.

We apply the Algorithm 2 based on the coarse resolution version of histogram with 20 bins. The results based on fine resolution version of histogram with 80 bins are rather similar, but a bit harder to visualize. The resultant heatmap is framed by a hierarchical clustering (HC) tree on its row axis, as reported on the panel (B) of Fig.\ref{fig:startspeed_ANOHT}. Percentages attached on each inter-node of the HC tree are calculated from 10,000 mimicked HC trees. We see via the HC tree branches in the order going from top to bottom that the branch-based-group of Hendrick's CH and Arrieta's CU, the group of Hendrick's CU and Dickey's KN, the group of Arrieta's FF and SI, and the group of Hendrick's SI and FC, are all confirmed as being authentic with significantly high probabilities. These memberships of the four pairs of pitcher-specific pitch-types are not mixing with any pitcher's pitch type outside of their groups.  Also the triplet branch-based-groups: Hendrick's CH, Arrieta's CU and Dickey's FF and the branch-based-group of five: Hendrick's FF, SI and FC and Arrieta's SL and CH, are also confirmed as being authentic as well. Therefore we may conclude that these two authentic branches: one on the slow end and one on the fast end of start-speed, are two large data-driven anchors of the tree geometry.

\begin{figure}[hbtp]
    \begin{center}
    	\includegraphics[scale=0.6]{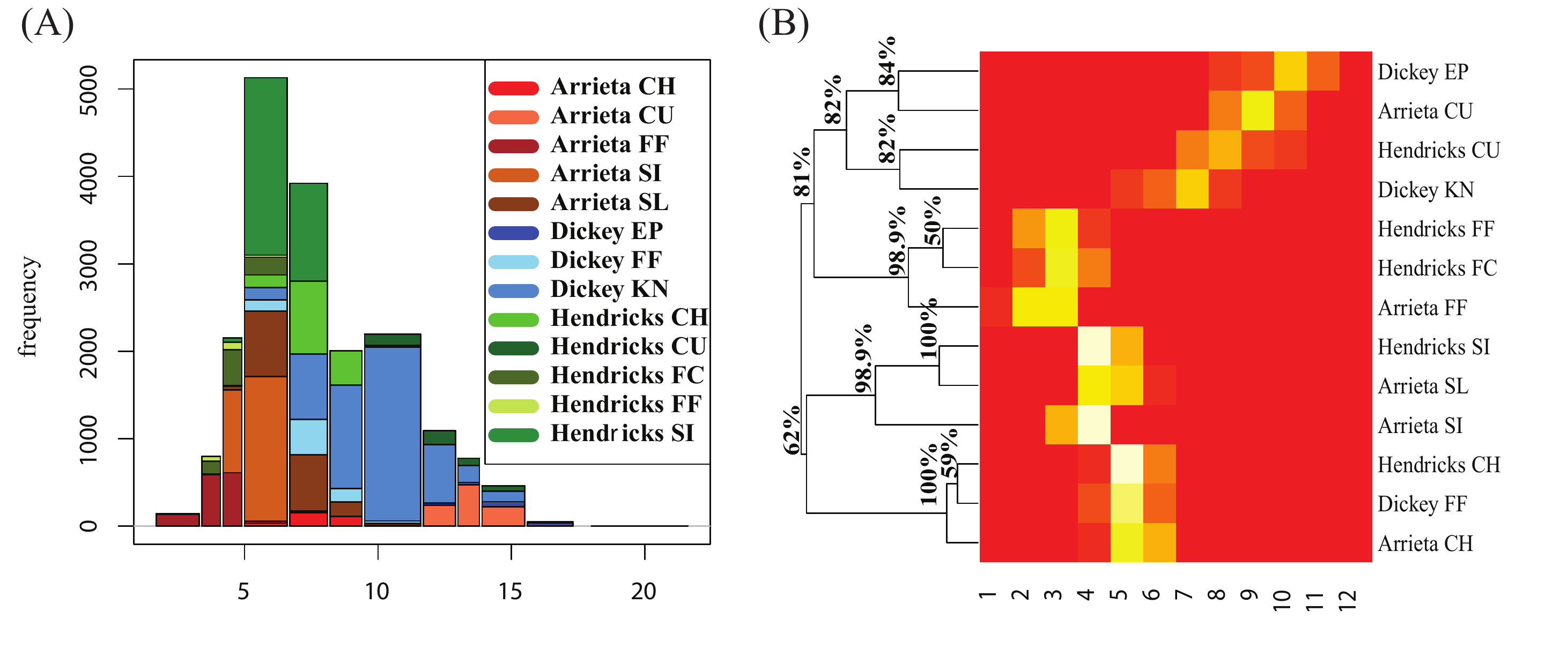}
    \end{center}
    \caption{The possibly gapped histogram in panel(A) for breaklength of pitches thrown by for Arrieta, Dickey and Hendricks in 2015-2016, color coded according to the 13 different pitching styles. Panel (B) shows the ANOHT tree percentages on the proportion matrix of 13 pitching styles.}
    \label{fig:breaklength_ANOHT}
  \end{figure}

Results from ANOHT on break-length are reported in two panels of Fig. \ref{fig:breaklength_ANOHT}. The histogram in panel (A) reveals that bins on the lower end are dominated by Arrieta and Hendrick's fastball, while bins on the larger end are dominated by Arrieta's CU and Dickey's KN. Also it is odd, but interesting to see that the bin located at 10 is nearly exclusively occupied by Dickey's KN. This exclusiveness implies that the Knuckleball-specific range of break-length is likely jointly achieved by the pitcher's unusual pitching mechanics and the unusual low start-speed as seen in bins on the lower end of histogram of start-speed in panel (A) of Fig. \ref{fig:hist_3pitcher}. This is another mechanistic pattern that can be possibly derived from ANOHT, but hardly could be easily derived from other methodologies.

For tree geometry among the 13 pitcher-specific pitch types, the HC tree superimposed on the row axis of heatmap reveals three authentic triplet branch-based-groups, as shown in the panel (B) of Fig. \ref{fig:breaklength_ANOHT}. These three triplet groups are: 1)Hendrick's FF and FC and Arrieta's FF; 2)Hendrick's SI and Arrieta's SL and SI; 3) Hendrick's CH and Dickey's FF and Arrieta's CH. All these three authentic branches are all on the lower end of break-length. In contrast, indexes pertaining to the branch-based-group memberships on the higher end are high, but not high enough to be comfortably confirmed as being authentic.

%[Fig.7 Here]

\section{Results on divorce data}
We then apply our ANOHT on a heavily censored  divorce dataset. One significant feature of such a dataset is that the heavily right-censoring typically causes many issues in parametric and semi-parametric inferences in survival analysis literature, see \cite{Padgett} and \cite{Bassiakos}. However it seems to have
relatively little effects on ANOHT. The key reason is that ANOHT can be based only on the reliable portion of histograms.

The dissolution of marriage dataset is based on a survey conducted in the U.S. studying 3,371 couples. The unit of observation is a couple and the event of interest is divorce. Couples lost to follow-up or widowed are treated as censored observations. We have three fixed covariates: education of the husband and two indicators of the couple ethnicity regarding whether the husband is black and whether the couple is mixed.

This dataset consists of only 1033 couples with completely observed event-time of dissolution of marriage in years, and 2338 couples with right censored event-times. The censoring rate is about $2/3$. The first factor is the years of husband's education with three categories: 1)coded as 0 for being less than 12 years; 2) coded as  1 for being between 12 to 15 years; 3) coded as 2 for being 16 or more years. The second factor is husband's ethnicity with two categories: 1) coded as 1 for the husband being black; 2) codes as 0 otherwise. The third factor is couple's ethnicity with two categories: 1) coded as 1 if the husband and wife have different ethnicity; 2)coded 0 otherwise. Censoring status as usual is coded as 1 for divorce and 0 for censoring (due to widowhood or lost to follow up).

Thus this dataset can be partitioned with respect to one single factor, or a pair of factors, or the three factors together. That is, the finest partition has 12 samples or treatments, in which each treatment is triple-coded. For instance the treatment `201' stands for the sample of couples with husband's education being more than 15 years (coded 2 in the first factor), non-black husband (coded 0 in the second factor) and having mixed ethnicity (coded 1 in the third factor).

\subsection{ANOHT on a divorce data}
\begin{figure}
\begin{center}

\includegraphics[height=5in, width=5.2in]{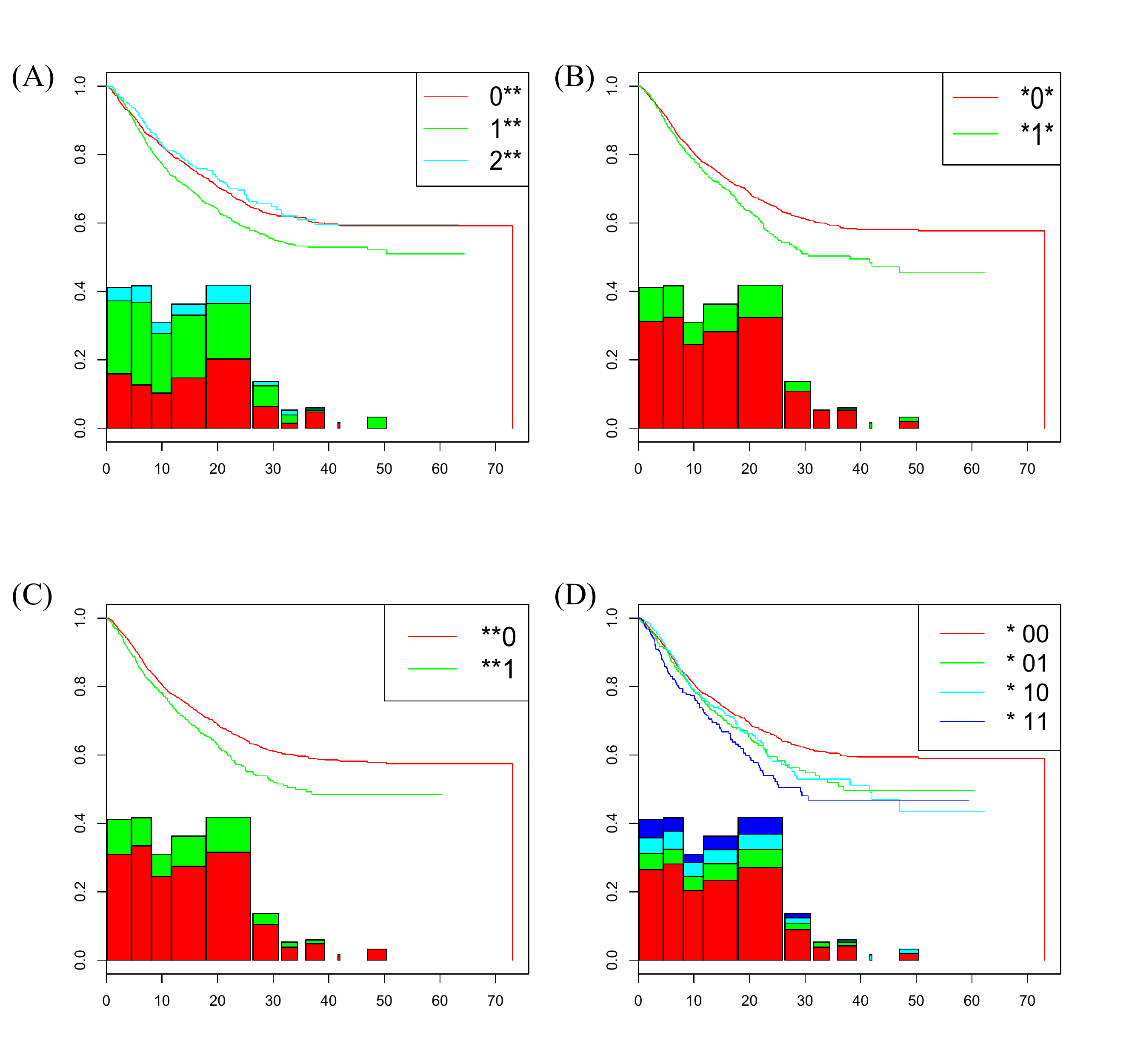}

\end{center}
\caption{The first phase of ANOHT results on divorce dataset: (A) three treatments via 1st factor; (B) 2nd factor; (C) 3rd factor; (D) four treatments with respect to the 2nd and 3rd factors.}
\label{fig:divorce1}
\end{figure}

The first phase of ANOHT on this divorce dataset on these four aspects are reported in the Fig. \ref{fig:divorce1}.
Overall the possibly-gapped histogram shows three evident gaps along its right tail. These gaps are not likely due to old age. Since they are followed by bins with significant heights, their locations seem interesting, but need more research attentions for pertinent interpretations. Specifically the panel (A) of Fig \ref{fig:divorce1} shows three Kaplan-Meier estimates of three survival functions with respect to three categories of education levels. It is rather interesting to note that the marriage with husbands having middle education level (1**) is likely to fail much earlier than marriages with husbands in either the lowest (0**) or highest (2**) education levels. The panels (B) and (C) of Fig. \ref{fig:divorce1} clearly reveal that the two categories: Black-husband (*1*) and mixed-ethnicity of couple (**1), respectively associate with lower survival rates. Further the category: Black-husband and mixed-ethnicity of couple (*11), has the lowest survival rate against the other three categories in panel (D) of Fig. \ref{fig:divorce1}. This survival rate is especially lower than the one belonging to the category: non-black husband and wife (*00).

The second phase of ANOHT on divorce dataset is reported in the figure below:

\begin{figure}[hbtp]
\begin{center}
	
	\includegraphics[scale=0.55]{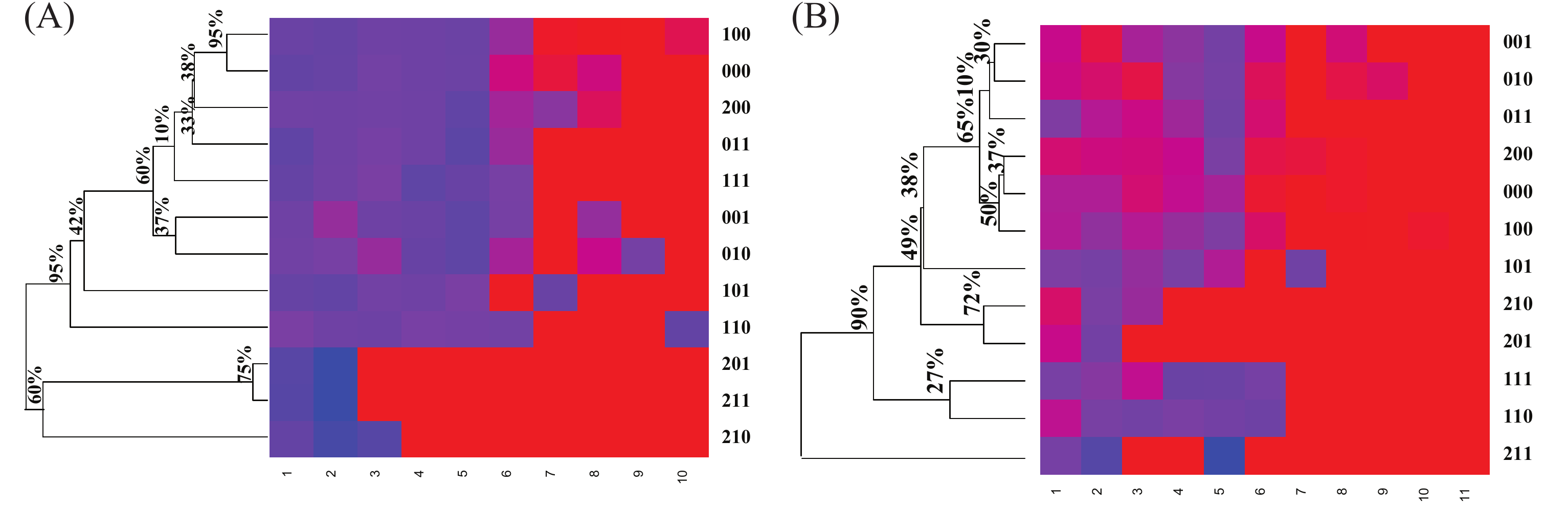}
	
\end{center}
	\caption{The second phase of ANOHT results on divorce dataset: (A) based on Kaplan-Meier survival function estimates on 12 categories; (B) based on the Nelson-Aalen cumulative hazard function estimates. }
	\label{fig:divorce2}
\end{figure}

The group of 6 treatments (categories): (000, 001, 010, 100, 011, 200), are seen in panels (A) and (B) of Fig. \ref{fig:divorce2}. They have similar Kaplan-Meier survival function and Nelson-Aalen cumulative hazard functions. Particularly its subgroup of 3 treatments:(000, 100, 200) shows very high similarity on both aspects among them. Such evidence is interpreted as that the factor of education has no effect on occurrence of event of marriage failure in non-black couple. In contrast, the presence of the subgroup of 3 treatments: (010, 001, 011), seems to imply that the occurrence of the event of marriage failure for a couple with husband having lowest education level is neither affected by the husband's ethnicity,  nor the couple's mixture of ethnicity.

From Kaplan-Meier survival function perspective, the tree on the heatmap of panel (A) of Fig. \ref{fig:divorce2} evidently shows one $95\%$ authenticity of the large branch of 9 treatments against the other branch of 3 treatments: (201, 211, 210). Another $95\%$ authenticity is found on the branch of two treatments: (100, 000). In contrast, from Nelson-Aalen cumulative hazard function perspective, there is only $90\%$ authenticity found on the outlier treatment: (211) against the rest of treatments.

\section{Conclusion}
In this note we demonstrate algorithmic data-driven computing for constructing a possibly-gapped histogram, and then develop two phases of Analysis of Histogram (ANOHT). Their practical values are clearly illustrated and manifested through Iris data, baseball pitching data and divorce data. Our data-driven computing simply and critically demonstrates interfaces of machine learning, information theory, and statistical physics. We believe such data analytic techniques would have very high potentials for very wide spectra of problems in sciences. 

It is surprising that a well-known Hierarchical clustering algorithm, which was first developed in 1950's for taxonomy (\cite{Sneath}), can help resolve a complex computational physical problem with exponential growth of complexity. Even more striking is that such a multi-scale physical problem explicitly contrasts and brings out the unspoken assumption of homogeneity implicitly assumed in all known statistical model selection techniques.

We believe that the existential issue of gap in a distribution will become more critical in this Big Data era ever than before. Its biological and mechanistic meanings should be a part of critical knowledge discovery in data-driven analysis. Particularly a possibly-gapped histogram can provide a fundamental new method of re-normalization on data via digital coding schemes. Such a data re-normalizing method in fact plays a critical role in unsupervised machine learning for knowledge discovery.

The merits of two phases of ANOHT are apparently seen in the three real examples, so are likely to be seen in many scientific researches. We have demonstrated that they have the capabilities to resolve issues never being discussed before, and at the same time to provide knowledge discoveries from wider perspectives. Further, since they are free from common, but unrealistic constraints and assumptions imposed by statistical modeling and analysis, their results should be more authentic and closer to reality.

Here we devote the rest of this section for implications of our developments on statistics. It is because of lacking continuity or smoothness assumptions in our developments, many types of potential complexity can be realistically embraced within our data analysis on 1D dataset. For instance, the multi-scale issue is clearly seen in the histograms of start-speed, as shown in Fig. \ref{fig:hist_arrieta}. This issue has to be resolved in a data-driven fashion because of the lack of prior knowledge. In sharp contrast, a histogram with pre-fixed number of equal-size bins is just not realistic. Here we emphasize that ignoring realistic information contents contained in data is hardly a right way of analyzing real data.

This multiscale feature also spells out the fact that all bins' boundaries are not global parameters. Thus the ensemble of candidate histograms is far from being fixed, but grows exponentially like the ensemble of all possible spin-configurations of Ising model in statistical physics. They render no uniform $1/\sqrt{n}$ or $1/n$ rates of convergence. Thus all statistical model selection techniques, such as AIC, BIC and minimum description length (MDL), in statistics are not applicable. The simple underlying technical reason is that all these model selection techniques require their ensemble of candidate models to be fixed and independent of $n$, so that the corresponding mathematical optimization problem can be well defined.

The implications of ANOHT on statistics are clearly demonstrated and contrasted through its applications on the Iris, MLB pitching and divorce data sets. The computing simplicity and informative results pertaining to the two phases of analyses via ANOHT come from the recognized interface of a physical problem and an unsupervised machine learning algorithm.

\vspace{0.5in}

\section*{\bf Ethics:} No ethical approvals are needed in this study.  All measurements pertaining to study subjects are available from two public websites.

\section*{\bf Data Accessibility:} The Iris data is available at UCI Machine Learning Repository:  \url{https://archive.ics.uci.edu/ml/datasets.html}). The pitching data is available in PITCHf/x database belonging to Major League Baseball via \url{http://gd2.mlb.com/components/game/mlb/}. The divorce data is available at: \url{http://data.princeton.edu/wws509/datasets/#divorce}

\section*{\bf Competing Interests:} We have no competing interests.

\section*{\bf Author's contributions:} H.F. designed the study. T.R. collected all data for analysis. H.F. and T.R analyzed the data. H.F and T.R. interpreted the results and wrote the manuscript. All authors gave final approval for publication.

\section*{\bf Funding:} No financial funding received for this study.

\section*{\bf Acknowledgement:} We thank Kevin Fujii for helps in pitching data retrieval.

%\bibliographystyle{acm.bst}
%\bibliography{sigproc2}

\end{document}